\documentclass[aps, prx,onecolumn,superscriptaddress,nofootinbib
]{revtex4-1}

\usepackage{graphicx}
\usepackage{amsmath}
\usepackage{setspace}
\usepackage{braket}
\usepackage{natbib,mathrsfs,accents}
\usepackage[dvipsnames]{xcolor}
\definecolor{myblue}{named}{MidnightBlue}
\definecolor{mygreen}{RGB}{0,120,0}
\usepackage[colorlinks=true,citecolor=myblue,linkcolor=myblue,urlcolor=myblue]{hyperref}
\usepackage{siunitx}
\usepackage[caption=false]{subfig}

\begin{document}
\pagenumbering{gobble}

\title{Topical White Paper: A Case for Quantum Memories in Space }
\author{Mustafa Gündoğan}
\thanks{Topical white paper submitted to Decadal Survey on Biological and Physical Sciences Research in Space 2023-2032. Primary author: mustafa.guendogan@physik.hu-berlin.de; +49 (030) 2093 82563}
\affiliation{Institüt für Physik, Humboldt Universität zu Berlin}
\author{Thomas Jennewein}
\affiliation{Institute for Quantum Computing and Department of Physics and Astronomy, University of Waterloo}
\author{Faezeh Kimiaee Asadi}
\affiliation{Institute for Quantum Science and Technology, and Department of Physics and Astronomy, University of Calgary}
\author{Elisa Da Ros}
\affiliation{Institüt für Physik, Humboldt Universität zu Berlin}
\author{Erhan Sağlamyürek}
\affiliation{Institute for Quantum Science and Technology, and Department of Physics and Astronomy, University of Calgary}
\affiliation{Department of Physics, University of Alberta}
\author{Daniel Oblak}
\affiliation{Institute for Quantum Science and Technology, and Department of Physics and Astronomy, University of Calgary}
\author{Tobias Vogl}
\affiliation{Institute of Applied Physics, Abbe Center of Photonics, Friedrich-Schiller-Universität Jena}
\affiliation{Cavendish Laboratory, University of Cambridge}
\author{Daniel Rieländer}
\affiliation{Fraunhofer Institute for Applied Optics and Precision Engineering}
\author{Jasminder Sidhu}
\affiliation{SUPA Department of Physics, University of Strathclyde}
\author{Samuele Grandi}
\affiliation{ICFO-Institut de Ciències Fotòniques, Barcelona Institute of Science and Technology}
\author{Luca Mazzarella}
\affiliation{Jet Propulsion Lab, California Institute of Technology}
\author{Julius Wallnöfer}
\affiliation{Institut für Theoretische Physik, Freie Universität Berlin}
\author{Patrick Ledingham}
\affiliation{Department of Physics and Astronomy, University of Southampton}
\author{Lindsay LeBlanc}
\affiliation{Department of Physics, University of Alberta}
\author{Margherita Mazzera}
\affiliation{Institute of Photonics and Quantum Sciences, SUPA,
Heriot-Watt University}
\author{Makan Mohageg}
\affiliation{Jet Propulsion Lab, California Institute of Technology}
\author{Janik Wolters}
\affiliation{Institute of Optical Sensor Systems, Deutsches Zentrum für Luft- und Raumfahrt e.V. (DLR)}
\affiliation{Institut für Optik und Atomare Physik, Technische Universität Berlin}
\author{Alexander Ling}
\affiliation{Centre for Quantum Technologies, National University of Singapore}
\author{Mete Atatüre}
\affiliation{Cavendish Laboratory, University of Cambridge}
\author{Hugues de Riedmatten}
\affiliation{ICFO-Institut de Ciències Fotòniques, Barcelona Institute of Science and Technology}
\author{Daniel Oi}
\affiliation{SUPA Department of Physics, University of Strathclyde}
\author{Christoph Simon}
\affiliation{Institute for Quantum Science and Technology, and Department of Physics and Astronomy, University of Calgary}
\author{Markus Krutzik}
\affiliation{Institüt für Physik, Humboldt Universität zu Berlin}

\begin{abstract}
    It has recently been theoretically shown that Quantum Memories (QM) could enable truly global quantum networking when deployed in space~\cite{Gundogan2021, Liorni2021} thereby surpassing the limited range of land-based quantum repeaters. Furthermore, QM in space could enable novel protocols and long-range entanglement and teleportation applications suitable for Deep-Space links and extended scenarios for fundamental physics tests.  In this white paper we will make the case for the importance of deploying QMs to space, and also discuss the major technical milestones and development stages that will need to be considered. 
\end{abstract}
\maketitle

\section{Introduction}
 \pagenumbering{arabic} 

A quantum memory (QM) is a device with the ability to store the quantum state of an incoming light pulse and release it on-demand at a later time~\cite{Afzelius2015}. QMs are required in several  protocols in quantum communication, quantum computing and optics. In the context of long-range links, the main function of QMs is to synchronize otherwise probabilistic two (or more)-photon transfer and detection events in order to speed up protocols, such as quantum repeaters~\cite{Briegel1998, Sangouard2011} and deterministic creation of multiphoton states~\cite{Nunn2013, Kaneda2017}. There are several figures of merit that characterize the performance of QMs:

\begin{itemize}
    \item Fidelity (F): the retrieved quantum state, $\rho'$, should be as close as possible to the input quantum state $\rho$. The fidelity is given by the overlap between these states, and can be calculated as: $\text{F}(\rho)=\text{Tr}\sqrt{\sqrt{\rho'}\rho\sqrt{\rho'}}$.
    \item Efficiency ($\eta$): The memory efficiency is the ratio of probability of detecting the output photon to that of the input photon\footnote{\textcolor{black}{ Although there has been progress towards space-based quantum communication with continuous variable protocols~\cite{Dequal2021} in this white paper we focus on discrete variables as they have a better loss tolerance and the fidelity of the final states do not necessarily depend on channel loss and memory efficiencies. }}; $\eta=\text{P}_{out}/\text{P}_{in}$. For ensemble based memories, it is usually limited by optical depth of the atomic ensemble and strength and temporal/spectral profile of the control pulses, mediating the storage.
    \item Storage time ($\tau$): in principle a QM should store the input quantum state as long as possible. This is usually limited by interatomic interactions, thermal effects, external magnetic or electric field noises and can be mitigated with several means. Today, QMs are pushing towards 1~s threshold~\cite{Yang2016}, while classical pulse storage for up to 1~h has been recently demonstrated~\cite{Ma2021_NC}. 
\end{itemize}

%In this white paper we would like to make the case for the importance of deploying QMs to space, and also discuss the major technical milestones that should be considered. 

\section{Quantum memories for space applications}
\label{sec:experiments}

It has recently been theoretically shown that QMs could enable truly global quantum networking when deployed in space~\cite{Gundogan2021, Liorni2021} thereby surpassing the limited range of land-based quantum repeaters. When combined with a set of tools such as high-fidelity entangled photon pair sources, entanglement swapping~\cite{Pan1998} and teleportation~\cite{Pirandola2015} schemes,  QMs will also enhance the distances for very-large scale entanglement and teleportation implementations in space, and enable novel applications and tests of gravity. Specifically, QMs could also find use in long-distance Bell tests~\cite{Rideout2012, Simon2003}  where they ensure the space-like separation of the detection events in order to close the locality loophole. This is achieved by delaying one of the pairs by a certain amount $\tau$ before the detection event. While short fibre-based delay lines ($\sim\mu$s) are sufficient for some experiments such as optical COW tests as envisioned in~\cite{Rideout2012}, QMs are definitely required in space-based experiments that require adjustable delay times of many milli-seconds to seconds, which is in particular the case for scenarios that involve large link distances such as for geostationary orbit to Earth (120~ms),  or Earth to Moon (1.3~s).  Long-lived QMs achieving high efficiencies ($\eta>70\%$ for $\tau\sim200$~ms) have already been demonstrated in the lab~\cite{Yang2016}. Hypothetically,  a fiber delay line for such a storage time\footnote{The best optical fibers have losses of around 0.14 dB/km at 1550~nm~\cite{Tamura2018} which translates into 0.028 dB/$\mu$s delay.  This makes it clear that any experiment that requires $\tau>1~$ms will have a transmission of less than $1\%$, thus making the fiber-based delay line extremely inefficient. Novel ZBLAN fibers are being projected to achieve $<0.01$~dB/km attenuation but they remain extremely rare and costly.} comes with a prohibitively high loss of $>5600$~dB.
% [TJ] here my loss estimate: length of fiber = 2e8 m/s * 200 ms = 40,000 km. 
% [TJ] losses = 40,000 km to * 0.14 dB/km = 5600 dB 

%In the following we overview the use of QMs in different space-based scenarios.   

\subsection{Ultra-long distance entanglement and teleportation experiments}

Long-range entanglement and teleportation experiments are fundamentally interesting for several reasons. Such experiments would open up the path towards long-distance quantum communication~\cite{Gundogan2021, Liorni2021, Boone2015,Sidhu2021}, all the way towards enabling tests of quantum mechanics across vast distances~\cite{Rideout2012, Simon2003}. Such experiments might also give hints towards the interplay between gravity and quantum physics \cite{Kiefer2012}.

Even for realizing a global network across Earth, using ground-based infrastructures alone will possibly be insufficient, and a hybrid quantum repeater network utilizing both space- and ground-based communication links may be the most effective solution, offering a large parameter space to investigate. In such a network, entanglement could be distributed over continental distances (up to a few thousand km) using ground fiber-based quantum repeaters~\cite{Sangouard2011, Vinay2017}, while longer distances could be reached using satellites, e.g. by performing entanglement swapping between quantum memories placed in LEO satellites~\cite{Gundogan2021, Liorni2021, Wallnofer2021}, or by entangling quantum memories located in separate  ground stations with entangled photon sources located on LEO satellites~\cite{Boone2015}. As is well established, optical fibers are commonly used to establish metropolitan quantum networks~\cite{valivarthi2016quantum, sun2016quantum}.  

Teleportation of unknown quantum states between spatially separated locations requires pre-distribution of entangled pairs as well as performing a Bell-state measurement. The result of the latter needs to be transmitted classically such that a correct unitary operation is applied to complete the teleportation protocol. It is, therefore, crucial to employ quantum memories that are capable of storing entangled states during the experiment. For example, teleportation without any post-selection requires a QM with storage times longer than the light's travel time between the measurement stations. For Earth-Moon distance this amounts to $\sim1.3$~s, as is considered for NASA's proposed Deep Space Quantum Link (DSQL)~\cite{Mazzarella2021, Jennewein2021}. It is within this context that space-based QMs could enable full teleportation that only performs a Bell-state measurement once all required signals are ready, and therefore circumvent the issue of post-selection. This would yield a drastic performance improvement over solutions without memory.

%QMs would find use in very-large scale teleportation experiments. A QM or a delay line is required to store the photon until the a classical signal from Alice is received for proper unitary transformation. A fibre-based delay line could be sufficient for short distances, hence short delays; however, for distances such as Earth-Moon separation a delay of $>1$~s is needed. It is thus clear that fibre-based delays are unpractical for such long distances as the associated optical losses are in the order of $\sim10^4$~dB. On the other hand there have been several light storage experiments (with bright pulses) with storage times $>1$~s have already been performed (see next sections) with losses between \textcolor{black}{10-40 dB}. 

\subsection{Long-distance matter-matter entanglement}

Tests of Bell’s inequality continue to be of great interest due to their connection to the foundations of quantum mechanics. An advantage of performing ultra-long-range Bell tests with space-like separated observers is examining the space-time features on quantum correlations and closing the locality and freedom-of-choice loopholes while experimenting~\cite{Hensen2015, Shalm2015,Giustina2015, Rosenfeld2017}. 
%In a more general case, to realize a loophole-free Bell test, we also need to address the question of the finishing time of a quantum measurement. 
Furthermore, it has been suggested that the measurement process is finished only once a gravity-induced state reduction takes place \cite{penrose1996gravity}. Therefore, testing ultralong-distance entanglement distribution where measurements are associated with the displacement of macroscopic masses and measuring the produced gravitational fields can improve our understanding of quantum gravity theories \cite{salart2008spacelike, kent2009proposed}.

A QM would be very helpful for such long-distance experiments in several ways. Utilizing quantum memories to store distributed photons before the detection events take place can provide the required time delay for addressing locality and freedom-of-choice loopholes. Besides, memories can significantly enhance the heralding efficiency when using human observers with free will~\cite{Bell1985, Kaltenbaek2004, cao2018bell}.

%Long-lived QMs would be helpful to create large-scale matter-matter entangled states in curved space-time.

%We may also consider entanglement of massive systems to further stretch the regimes in which quantum mechanics has been tested. [Proposal by Kent for Bell Tests with moving large masses. Gravitational collapse models, go beyond superposition to see if entanglement of massive bodies can further constrain alternative theories.]

\section{Physical Quantum Memory Candidate Systems}

While many different QM systems are under development, we primarily consider ensemble-based QM systems which we deem suitable to satisfy the requirements of the space-based experiments mentioned above. These systems allow input-output type memory operation in a way that they can be interfaced with external single photon or entangled photon pair sources. Furthermore, they can possess important features including multi-mode storage capacity, large spectral bandwidth and being more robust to small beam misalignments. The following systems\footnote{Although there is a great progress in quantum networking with single color centers~\cite{Bhaskar2020,Pompili2021}, we exclude them in this white paper as memory platforms due to possible incompatibility with the DSQL.} have thus the potential of being integrated into NASA's planned DSQL.

\subsection{Warm vapour memories}
\noindent 
Optical storage with room temperature alkali atoms are particularly interesting since it relies neither on complex laser trapping techniques nor on cryogenic cooling. This makes them promising candidates for field operation such as under sea or in space. Recent years saw significant progress in warm vapor memories. These include development of noiseless, fast memories~\cite{Kaczmarek2018, Finkelstein2018}, temporal multimode storage~\cite{Hosseini2011, Main2021}$, \sim$GHz-bandwidth operation~\cite{Wolters2017, Main2021} and highly efficient ($\eta>80\%$) storage with fidelities above the no-cloning limit~\cite{Guo2019}. Furthermore, record storage time of $\sim1$~s has been observed with storage of bright pulses~\cite{Katz2018} by operating at the so-called `spin exchange relaxation-free' regime. The storage time in these systems has the potential to be pushed beyond hours by transferring the spin excitation from alkali atoms to noble-gas nuclear spins via spin exchange collisions~\cite{Katz2021}. 

Amongst others, compact warm vapor-based optical spectroscopy setups have been operated as optical frequency references on sounding rocket missions \cite{Dinkelaker2016, Lezius2016,Doringshoff2019}, demonstrating the maturity of diode laser systems and related vapor cell based technologies. Further integration towards cubesat form factor compatible, integrated atomic systems for a frequency reference application has been demonstrated without reduction of performance \cite{Strangfeld2021}.

\subsection{Laser-cooled atomic systems}
\noindent 
Laser-cooled atomic systems are well-established platforms for quantum information storage. 
Even though the setups required to laser cool atomic ensembles are relatively more complex than the ones implemented for warm vapours, the lower temperatures of a cold atom cloud inhibits thermal diffusion and grants long coherence times. Of particular interest is photon storage in ultracold quantum gases such as Bose-Einstein condensates~\cite{Zhang2009,Riedl2012,Saglamyurek2021}: not only does the reduction of thermal motion offer long storage times, but the large optical depth allows for high write-read efficiency and the refined state preparation increases the fidelity of the storage in the atomic spin state.
In the past years optical memory has been demonstrated in these systems achieving high efficiencies~\cite{Yang2016,Cho2016}, long storage time~\cite{Dudin2013}, and  temporal~\cite{Cho2016, Heller2020} and spatial multiplexing~\cite{Lan2009,Pu2017}.

In parallel to these efforts, the past decade has witnessed extensive research and technological development to realize cold-atoms based setups for space applications. Deployment of these experiments in a microgravity environment would enable both fundamental research and promise enhanced performance for practical applications in navigation and earth observation.
In this context, generation and coherent manipulation of cold atom ensembles and Bose-Einstein condensates have already been demonstrated in microgravity environments including orbiting platforms such as parabolic flights~\cite{Langlois2018, Barrett2016}, drop-towers~\cite{Muntinga2013,Deppner2021}, sounding rockets~\cite{Becker2018,Lachmann2021}, and now even on Tiangong-2~\cite{Liu2018} and the ISS~\cite{Aveline2020}. Further launches to ISS are planned such as the ACES/PHARAO atomic clock~\cite{Laurent2015}, or the dual-species atomic experiment facility BECCAL~\cite{Frye2021}. Various proposals are in development for deployment on cubesats such as CASPA~\cite{Devani2020} or as part of large scale missions such as AEDGE~\cite{Aedge2020}. Very similar systems and infrastructures can lay the foundations of future space-based quantum memories for deployment in orbit.

\subsection{Rare-earth ion doped crystals (REIDs)}
Rare earth ions possess exceptionally long coherence times (narrow homogeneous linewidths) both on optical and spin transitions at cryogenic temperatures. In particular, their narrow optical transitions are directly linked to their unique electronic configurations, in which their active $4f$ orbitals are embedded within the filled outer shells, providing protection from external perturbations. This feature, in conjunction  with the absence of thermal motion, render REIDs as a high-performance QM platform. The recent achievements include but are not limited to; heralded entanglement generation between two QMs~\cite{Lago-Rivera2021, Liu2021} in a quantum repeater setting, bright pulse storage from minute~\cite{Heinze2013} to hour-long time scales~\cite{Ma2021_NC} and demonstration of temporal~\cite{Usmani2010,Gundogan2013,Seri2019} spectral~\cite{Sinclair2014,Saglamyurek2016} and spatial~\cite{Yang2018} multimode storage. The other research direction is the miniaturization of these devices: waveguide structures~\cite{Saglamyurek2011, Marzban2015, Corrielli2016} and nanophotonic cavities~\cite{Zhong2017, Dibos2018} offer an enhanced compactness and interaction strength. The narrow hyperfine level separation usually limits the storage bandwidth of the on-demand operation to a few MHz~\cite{Jobez2015,Gundogan2015}; however, the recently demonstrated hybridized electronic-nuclear spin levels in highly anisotropic host materials could enable the storage of large bandwidth photons while retaining long coherence times.~\cite{Businger2020}.  

In this way, REIDs can fulfill the requirements of experiments outlined in Sec.~\ref{sec:experiments} by combining  compactness with efficient and  long-lived storage capability for broadband pulses. They are thus among the promising candidates for space applications with the development of miniature, space-compatible cryostats~\cite{You2018}.

\section{Memory-compatible photon sources} 

Photon sources intended to interface with QM systems should match QM-specific bandwidth and wavelength for  optimum coupling. They must also have small footprints and robust architecture for operation in space. We consider two types of single photon sources that are both memory-compatible and suitable for operation in a space environment.

\subsection{SPDC sources}

Spontaneous parametric down-conversion (SPDC) has been the most established method for creating entangled photon pairs for more than three decades~\cite{Kwiat1993}. It relies on a nonlinear crystal being pumped with a strong laser beam which creates correlated photon pairs in different degrees of freedoms, e.g. polarization, angular momentum or frequency. They operate at room temperature, and have already been deployed in space. In addition to the entangled photon pair source on board the MICIUS satellite~\cite{Yin2020}, a miniaturized SPDC source  on board a cubesat successfully demonstrated entangled state generation~\cite{Villar2020}. Notably, the robustness of such space-compatible systems was proven by an SPDC source that survived a rocket explosion during its launch without showing a significant reduction of performance~\cite{Tang2016}. 

By itself, the emission from  SPDC is typically wideband and special techniques are required in order to achieve spectral matching of the SPDC photons with a QM.  The SPDC output can be filtered to reduce the photon bandwidth; however, this significantly reduces the brightness of the source. More effectively, the SPDC source can be placed inside a resonant cavity to reduce the linewidth while maintaining a high level of photon counts, which can yield very narrow, tunable bandwidth. For example, a highly degenerate photon-pair source that emits one photon resonant with a Pr-doped REID QM and the other at the telecommunications C-band~\cite{Fekete2013} was successfully used for entangling two crystals~\cite{Lago-Rivera2021}, and a sub-MHz linewidth source that operates at the Rb D1 line~\cite{Rambach2016} was demonstrated. Even though there is great progress towards creating such narrow band sources, the strict bandwidth requirements can further be relaxed for coupling to larger bandwidth QMs.

\subsection{Solid-state single photon sources}
Single photon emitters embedded in solid-state media~\cite{Aharonovich2016, Atature2018} are another alternatives to entangled photon pair sources for long-range entangled state creation~\cite{Sangouard2007}. A single photon pulse is sent to a beam splitter and thus a path-entangled state is created which can then be used to entangle two distant QMs~\cite{Usmani2012}. 
Currently the best single photon sources in terms of spectral purity and brightness are semiconductor quantum dots (QDs)~\cite{Senellart2017} that need cryogenic cooling. Even though on-demand storage and retrieval of QD single photons has been proved challenging due to the spectral mismatch between memory and  single photons, \textcolor{black}{generation of sub-natural linewidth photons from QDs~\cite{Matthiesen2012} may become crucial towards achieving this goal.} There have also been recent experiments that demonstrated coupling between QD single photons and atomic ensembles in the form of slow light~\cite{Siverns2019} or storage with a predetermined delay time~\cite{Tang2015}. In parallel to these efforts, there have been recent work towards the realization of a QM based on the nuclear spin ensemble in semiconductor QDs~\cite{Denning2019, Gangloff2019}. Such a system would finally realize a perfect single photon source with a built-in QM capability. 

\textcolor{black}{Defect centers in diamond are one of the most promising single photon sources linked to a local spin-based memory. The nitrogen-vacancy (NV) center is the most established spin-photon interface with impressive spin, but modest optical properties due to its sensitivity to electric noise and weak zero-phonon line (ZPL) efficiency. Recently emerging alternative to the NV center is the family of group-IV color centers, such as SiV, GeV and SnV, which are insensitive to the environmental noise due to their inversion symmetry~\cite{Hepp2014}, even when incorporated into nanophotonic structures~\cite{Sipahigil2016, Sohn2018, Trusheim2020}. The resulting stable emission line, combined with high ZPL efficiency, transform-limited linewidths down to $\sim30$~MHz~\cite{Trusheim2020} with coherent access to electronic spins~\cite{Debroux2021} at cryogenic temperatures makes these color centers a perfect candidate as single photon sources with built-in QM capability~\cite{Sipahigil2017}}.

Finally, single photon emitters in 2D materials~\cite{Toth2019} are nascent alternatives to the mature semiconductor QDs and diamond defects. Among these, emitters in hexagonal boron nitride (hBN) have further advantage of a large coverage of emission lines from the UV up to the NIR~\cite{Bourrellier2016,Camphausen2020}. Although they usually have large linewidth at room temperature, placing them in optical microcavities~\cite{Vogl2019} has proven to be an efficient way to drastically improve the spectral purity and linewidth. Finally, the recent observation of transform limited ($<100$~MHz) single photons~\cite{Dietrich2020} from hBn defects \textcolor{black}{could} bring coupling of these photons to QMs one step closer to reality \textcolor{black}{if the observed spectral diffusion can be mitigated}. In fact, these hBN-based emitters have already been qualified for space applications~\cite{Vogl2019b}. 

\section{Critical Technologies and Mission Demonstrations}

Deployment of QMs in space will require a sequence of milestones and demonstrations before such a mission can be realized. We should note that the different quantum memory platforms we summarized above are all in different stages of development and they offer different performances for different applications. For example, the current status of single photon level experiments with REIDs are relatively advanced~\cite{Seri2017,Jobez2016, Lago-Rivera2021} but these systems require cryogenics and even high B-fields for specific experiments ($>1$~T)~\cite{Zhang2015,Ma2021_NC}. On the other hand, such experiments are not nearly as advanced in warm vapour systems; nevertheless they do not need cryogenics or additional trapping lasers such as those needed in ultracold atoms, thus opening a path towards miniaturization and robust integration into satellite platforms. Along these lines a space compatible, miniaturized frequency references based on warm Rb vapour have recently been demonstrated~\cite{Strangfeld2021}. Converting such a setup for an optical memory seems feasible upon the integration of modulators and other laser sources in the very near future and, thus, it is reasonable to expect the first demonstration of space-based optical storage experiments sometime in the next five years.

%The deployment of QMs in space will require a sequence of milestones and demonstrations before such a mission can be realized. We should note that the different quantum memory platforms we summarized above are all in different stages of development and they offer different performances for different applications. For example, temporal multiplexing is extremely important for effective quantum communication; however, the REIDs which can possess high temporal multimode capacity in a single memory require cryogenics. On the other hand, long-lived QMs based on warm vapours usually require multiple memory cells or spatial modes for temporal multiplexing~\cite{Nunn2013} and with some protocols they can store t but they do not need cryogenics or additional trapping lasers, thus opening a path towards miniaturization and robust integration into satellite platforms. Along these lines a space compatible, miniaturized frequency references based on warm Rb vapour have recently been demonstrated~\cite{Strangfeld2021}. Converting such a setup for an optical memory seems feasible upon the integration of modulators and other laser sources in the very near future and, thus, it is reasonable to expect the first demonstration of space-based optical storage experiments sometime in the next five years.
%\footnote{In principle, there are some experimental works that demonstrate temporal multimode operation with warm vapours~\cite{Main2021}, but they do not utilize the long-lived ground state storage that is required for the DSQL.} 
\subsection{Technological and experimental milestones towards QMs in space}
Below we summarize some of the major experimental and technical milestones that we anticipate on the path towards the deployment of a fully functional QMs enhanced link in space:
\begin{itemize}
    \item {\bf Photon Sources: }Demonstrate interfacing of the QM system with single-photon sources. This necessitates that the source can be implemented with sufficient photon extraction and repetition rates, as well as the  required level of control over spatial, temporal, spectral modes.
    
    \item {\bf Miniaturized and rugged QM platforms: }Miniaturization and integration techniques to successfully develop small-form-factor and space-compatible quantum memories.
    
    \item {\bf Free-space Link:} Coupling of the photons travelling over a fluctuating and turbulent channel to the spatial modes required by a QM. This could be either coupling to a single mode beam, or a `few mode beam'. This may include both active and passive manipulation of the optical beam and wave-fronts. 
    
    \item {\bf QM operation and synchronisation:} Study aspects of operations, timing synchronisation and reference frame alignment required to enable operations of the QM, in particular using the   photon sources and the free-space link technology previously mentioned.
    
    \item {\bf Bell-state measurement: } As an important tool for a quantum entanglement distribution or teleportation protocol, the entangling feature of two-photon Hong-Ou-Mandel interference (HOM) must be demonstrated when fluctuating free-space links are considered.  The demonstration of HOM interference for true single photons emitted by separate QMs sent over free-space channels must be shown.
    
    \item {\bf Satellite to ground link: }  Storage of single photons transmitted from a satellite-based photon source to  a QM located in a ground station will demonstrate the viability of the scheme. Wave-front shaping techniques will have to be utilized to achieve good coupling of the distorted free-space beam to the QM. 
    
    \item{\bf Supporting technology:} Operating and manipulating QM in space will also necessitate the development of space compatible lasers, driver electronics and vacuum systems. In addition to wavelengths that were already operated in space (such as Rb wavelengths at 780 nm and 795 nm) space-compatible lasers at various other quantum memory wavelengths (580 nm, 606 nm, 894 nm, 980 nm, 1530 nm, etc.) should be developed. Another important technology will be miniaturized vacuum and cryogenic systems. Lastly, efficient quantum frequency conversion systems will be needed for the conversion between memory wavelengths and the telecommunication wavelengths that are commonly used in fibre infrastructres.
    
\end{itemize}

\subsection{Mission scenarios }
Here we summarize a possible path towards the realization of the full potential of QMs in space.

\begin{itemize}
   \item {\bf Proof-of-principle demonstration of in-orbit memory operation:} The first in-orbit memory experiments should demonstrate the robustness and rigidity of these systems and their ability to operate in space environment. 
   
   \item {\bf Photon-matter entanglement between ground and space:} By implementing Ground to Space, or Space to Ground, transmission and storage of  photons from an entangled photon source will demonstrate  entanglement of a `flying' qubit with the matter QM. While only one link is utilized, a long-distance test of entanglement is possible.  
   
  \item {\bf QM assisted quantum communication:} With the simultaneous loading of two QM with signals sent between two separate ground stations and the satellite, QM assisted quantum communication is achieved in a way explained as downlink or uplink scenarios in Ref.~\cite{Gundogan2021} by using protocols in Refs.~\cite{Panayi2014, Luong2016}. Demonstrating such a memory advantage~\cite{Bhaskar2020, Langenfeld2021} in space is an important first step towards a scalable global quantum network, and is expected to  outperform the direct, simultaneous transmission of both entangled photons.
  
\end{itemize}

\section{Conclusion}
The recent advances of QM based on ensemble systems have enabled several important scenarios and implementations of space based quantum networks, that were previously considered infeasible. As an example, NASA's planned DSQL could significantly benefit from the deployment of QMs in space. It is thus critical to develop space compatible QM systems and the required R$\&$D road map for development of key technologies.

\pagebreak

\bibliography{Bib}
\acknowledgments
This work was made possible by funding from the European Union's Horizon 2020 research and innovation programme under the Marie Skłodowska-Curie grant agreement No.~894590 (M.G.) and the support from the German Aerospace Center (DLR) through funds provided by BMWi (OPTIMO, No.~50WM1958 and OPTIMO-II, No.~50WM2055) (M.G and M.K). A portion of this research was carried out at the Jet Propulsion Laboratory, California Institute of Technology, under a contract with the National Aeronautics and Space Administration (80NM0018D0004). 

\end{document}